\documentclass[fleqn,usenatbib]{mnras}

\usepackage[table]{xcolor}
\usepackage{longtable}
\usepackage{pdflscape}
\usepackage[maxfloats=110]{morefloats}

\usepackage[T1]{fontenc}

\DeclareRobustCommand{\VAN}[3]{#2}
\let\VANthebibliography\thebibliography
\def\thebibliography{\DeclareRobustCommand{\VAN}[3]{##3}\VANthebibliography}

\usepackage{graphicx}	
\usepackage{amsmath}	
\usepackage{amssymb}	
\usepackage{xspace}

\usepackage{newtxtext,newtxmath}

\title[BEBOP III: Kepler-16]{BEBOP III. Observations and an independent mass measurement of  Kepler-16~(AB)~b -- the first circumbinary planet detected with radial velocities\thanks{Based on observations collected at the Observatoire de Haute-Provence (CNRS, France)}}

\author[A.H.M.J. Triaud et al.]{Amaury H.M.J. Triaud$^{1}$\thanks{E-mail: a.triaud@bham.ac.uk}, %
Matthew R. Standing$^{1}$, %
Neda Heidari$^{2,3,4}$, %
David V. Martin$^{5,6}$, %
Isabelle Boisse$^{2}$, \newauthor%
Alexandre Santerne$^{2}$, %
Alexandre C.~M. Correia$^{7,8}$, %
Lorena Acu\~na$^{2}$, %
Matthew Battley$^{9,10}$, %
Xavier Bonfils$^{11}$,\newauthor %
Andr\'es Carmona$^{11}$, %
Andrew Collier Cameron$^{12}$, %
P\'ia Cort\'es-Zuleta$^{2}$, %
Georgina Dransfield$^{1}$, %
Shweta Dalal$^{13}$, \newauthor %
Magali Deleuil$^{2}$, %
Xavier Delfosse$^{11}$, %
Jo\~ao Faria$^{14,15}$,
Thierry Forveille$^{11}$, %
Nathan C. Hara$^{16}$, \newauthor %
Guillaume H\'ebrard$^{13}$, %
Sergio Hoyer$^{2}$, %
Flavien Kiefer$^{13}$, %
Vedad Kunovac$^{1}$, %
Pierre F.~L. Maxted$^{17}$, \newauthor %
Eder Martioli$^{13,18}$, %
Nikki Miller$^{17}$, %
Richard P. Nelson$^{19}$, %
Mathilde Poveda$^{20,21}$, \newauthor %
Hanno Rein$^{22}$, %
Lalitha Sairam$^{1}$, %
St\'ephane Udry$^{16}$, %
Emma Willett$^{1}$\\
$^{1}$School of Physics \& Astronomy, University of Birmingham, Edgbaston, Birmingham, B15 2TT, UK\\
$^{2}$Aix Marseille Univ, CNRS, CNES, LAM, Marseille, France\\
$^{3}$Department of Physics, Shahid Beheshti University, Tehran, Iran\\
$^{4}$Laboratoire J.-L. Lagrange, Observatoire de la C\^ote d’Azur, Universit\'e de Nice-Sophia Antipolis, CNRS, Campus Val-rose, 06108 Nice Cedex 2, France\\
$^{5}$Department of Astronomy, The Ohio State University, 4055 McPherson Laboratory, Columbus, OH 43210, USA\\
$^{6}$NASA Sagan Fellow\\
$^{7}$CFisUC, Departamento de F\'isica, Universidade de Coimbra, 3004-516 Coimbra, Portugal\\
$^{8}$IMCCE, UMR8028 CNRS, Observatoire de Paris, PSL Universit\'e, 77 av. Denfert-Rochereau, 75014 Paris, France\\
$^{9}$Department of Physics, University of Warwick, Gibbet Hill Road, Coventry CV4 7AL, UK\\
$^{10}$Centre for Exoplanets and Habitability, University of Warwick, Gibbet Hill Road, Coventry CV4 7AL, UK\\
$^{11}$Universit\'e Grenoble Alpes, IPAG, 38000, Grenoble, France; CNRS, IPAG, 38000, Grenoble\\
$^{12}$Centre for Exoplanet Science / SUPA, School of Physics and Astronomy, University of St Andrews, North Haugh, St Andrews, Fife, KY16 9SS, UK\\
$^{13}$Institut d’Astrophysique de Paris, UMR7095 CNRS, Universit\'e Pierre \& Marie Curie, 98bis boulevard Arago, 75014 Paris, France\\
$^{14}$Instituto de Astrof\'iısica e Ci\^encias do Espa\c co, Universidade do Porto, CAUP, Rua das Estrelas, 4150-762 Porto, Portugal\\
$^{15}$Departamento de F\'isica e Astronomia, Faculdade de Ci\^encias, Universidade do Porto, Rua do Campo Alegre, 4169-007 Porto, Portugal\\
$^{16}$Observatoire Astronomique de l’Universit\'e de Gen\`eve, Chemin de Pegasi 51, 1290 Versoix, Switzerland\\
$^{17}$Astrophysics Group, Keele University, ST5 5BG, UK\\
$^{18}$Laborat\'{o}rio Nacional de Astrof\'{i}sica, Rua Estados Unidos 154, 37504-364, Itajub\'{a} - MG, Brazil \\
$^{19}$Astronomy Unit, Queen Mary University of London, Mile End Road, London, E 14NS, UK\\
$^{20}$Universit\'e Paris Est Cr\'eteil and Universit\'e de Paris, CNRS, LISA, 94010 Cr\'eteil, France\\
$^{21}$Universit\'e Paris-Saclay, UVSQ, CNRS, CEA, Maison de la Simulation, 91191, Gif-sur-Yvette, France\\
$^{22}$Department of Physical and Environmental
Sciences, University of Toronto at Scarborough, Canada\\
}

\date{Accepted XXX. Received YYY; in original form ZZZ}

\pubyear{2021}

\begin{document}

\label{firstpage}
\pagerange{\pageref{firstpage}--\pageref{lastpage}}
\maketitle

\begin{abstract}
The radial velocity method is amongst the most robust and most established means of detecting exoplanets. Yet, it has so far failed to detect circumbinary planets despite their relatively high occurrence rates. 
Here, we report velocimetric measurements of Kepler-16A, obtained with the SOPHIE spectrograph, at the Observatoire de Haute-Provence's 193cm telescope, collected during the BEBOP survey for circumbinary planets. Our measurements mark the first radial velocity detection of a circumbinary planet, independently determining the mass of Kepler-16~(AB)~b to be $0.313 \pm 0.039\,{\rm M}_{\rm Jup}$, a value in agreement with eclipse timing variations. Our observations demonstrate the capability to achieve photon-noise precision and accuracy on single-lined binaries, with our final precision reaching $\rm 1.5~m\,s^{-1}$ on the binary and planetary signals. Our analysis paves the way for more circumbinary planet detections using radial velocities which will increase the relatively small sample of currently known systems to statistically relevant numbers, using a method that also provides weaker detection biases. Our data also contain a long-term radial velocity signal, which we associate with the magnetic cycle of the primary star. 

\end{abstract}




\section{Introduction}
Circumbinary planets are planets that orbit around both stars of a binary star system. Long postulated \citep{Borucki:1984,Schneider:1994lr}, the first unambiguous discovery of a circumbinary planet came with Kepler-16 \citep{Doyle:2011vn}, detected by identifying three transits within the lightcurve of an eclipsing binary system monitored by NASA's {\it Kepler} mission \citep{Borucki:2011qy}. {\it Kepler} went on to detect another 13 transiting circumbinary planets, in 11 systems \citep{Martin2018,Socia:2020}, with another two systems found using {\it TESS} \citep{Kostov2020,Kostov2021}. A number of circumbinary planets systems are suspected from eclipse timing variations of binaries on the main sequence \citep[e.g.][]{Borkovits2016,Getley2017}, and stellar remnants \citep[e.g.][]{Marsh:2013lr,Han2017} but most are disputed \citep[e.g.][]{Mustill2013}, and some disproven \citep[e.g.][]{Hardy2015}. Other detections include HD~106906~b, in direct imaging \citep[][]{Bailey2014} and OGLE-2007-BLG-349L(AB)c with the microlensing method \citep{Bennett:2016}. 

Despite successes with almost every observational methods, no circumbinary planet signal has been detected using radial velocities yet. In addition, radial velocities have detected many planets with masses compatible with currently known circumbinary planets. This is remarkable since radial velocities are one of the earliest, most established and efficient method of exoplanet detection. The system closest to a circumbinary configuration identified thus far is HD~202206 (\citealt{Correia:2005lr}, see Sect.~\ref{sec:discussion}).

The radial velocity method has a number of advantages over the transit method. First,  it is less restrictive in term of the planet's orbital inclination thus providing a weaker bias towards short orbital periods. Furthermore, the signal can be obtained at every orbital phase, and the method is more cost-effective and easier to use over a longer term thanks to using  ground-based telescopes \citep[][]{Martin2019}. Additionally, radial velocities provide the planet's mass, its most fundamental parameter. While the transit method can provide a mass when eclipse timing variations are detected, most circumbinary exoplanets unfortunately remain without a robust mass determination with eclipse timing variations mostly providing upper limits \citep[e.g.][]{Orosz2012,Schwamb:2013kx,Kostov2020}. Only four of the known circumbinary planets have eclipse-timing mass estimates inconsistent with 0 at $>3\sigma$. 
The present and future {\it TESS} and {\it PLATO} missions \citep{Ricker:2014qv, Rauer:2014} are set to identify several more transiting circumbinary planet candidates \citep[e.g.][]{Kostov2020,Kostov2021}. However, these are unlikely to produce many reliable mass measurements, in good part due to rather short observational timespans compared to {\it Kepler}'s.

Overall, radial velocities are essential to create a sample of circumbinary planets that is both greater in number and less biased than the transit sample. This will allow a deeper understanding of circumbinary planets: their occurrence rate \citep[][]{Martin:2014lr,Armstrong:2014fk}, multiplicity \citep{Sutherland:2019,Orosz:2019}, formation and evolution \citep[e.g.][]{Chachan2019,Pierens2020, Penzlin2021}, and dependence on binary properties \citep{Munoz:2015,Martin:2016,Li:2016,Martin:2019b}.

In 2017 we created the BEBOP survey \citep[Binaries Escorted By Orbiting Planets;][]{Martin2019}, as a blind radial velocity survey for circumbinary planets. Prior to this, the most extensive radial velocity effort had been produced by the TATOOINE survey \citep[][]{Konacki:2009lr}, but the survey unfortunately did not yield any discoveries. One issue likely affected the survey, from which BEBOP learnt a great deal: TATOOINE targeted double-lined binaries, which was logical. Double-lined binaries are brighter, both stars can have model-independent mass measurements, and one can in principle measure the Doppler displacement caused by a planet on each of the two components. However, disentangling both components from their combined spectrum accurately is a complex task, and despite photon noise uncertainties regularly reaching 2 to $4\,{\rm m\,s}^{-1}$, the survey returned a scatter of order 15 to $20 \,{\rm m\,s}^{-1}$ \citep[][]{Konacki:2009lr,Konacki:2010lr}. Indeed, \citet[][]{Konacki:2010lr} recommended single-lined binaries as a solution, but too few were known at the time. Since, radial velocities have been used to constrain the binary parameters, which helps refining the planetary parameters, but not to search for circumbinary planets themselves \citep[e.g.][]{Kostov2013,Kostov2014}.

Thanks to the advent of exoplanet transit experiments, increasing amounts of low-mass, single-lined eclipsing binaries are being identified \citep[e.g.][]{Triaud:2013lr,Triaud:2017yu,Boetticher2019,Lendl2020,Mireles2020,Acton2020}. The BEBOP survey was constructed solely using sufficiently faint secondaries, to avoid detection with spectrographs, such that we could in principle reach a radial velocity precision comparable to that around single stars of the same brightness. In principle, this ought to provide an accuracy of order ${1~\rm m\,s}^{-1}$. Our survey is ongoing, and uses the CORALIE, SOPHIE, HARPS, and ESPRESSO spectrographs. Preliminary results were published in \citet{Martin2019} (BEBOP I). In \citet[][under review, BEBOP II]{Standing2021}, we describe our observational protocol, the methods we use to detect planets, as well as how we produce detection limits.

In this paper we detail a complementary project to BEBOP's blind search. Between 2016 and 2021 we monitored Kepler-16, a relatively bright (Vmag = 12) single-lined eclipsing binary system with a primary mass $M_{\rm 1} = 0.65~\rm M_\odot$ (a K dwarf), a secondary mass $M_{\rm 2} = 0.20~\rm M_\odot$ (a mid M dwarf), and an orbital period $P_{\rm bin} = 41.1~\rm days$. The system is $75~\rm pc$ distant and known to host a circumbinary gas giant planet with a mass $m_{\rm pl} = 0.33~\rm M_{\rm Jup}$, and a period $P_{\rm pl} = 229~\rm days$. Our observations demonstrate that we can indeed recover the Doppler reflex signature of a circumbinary planet. Our results act to both validate and assist our broader search for new planets. Furthermore, we can derive a ``traditional'' Doppler mass measurement for the planet, to be compared with that derived from photometric eclipse and transit timings. Finally, our long baseline is sensitive to additional planets, in particular to any that would occupy an orbit misaligned to the transiting inner planet's.

\section{Velocimetric observations on Kepler-16}

Between 2016-07-08 and 2021-06-23, we collected 143 spectra using the high-resolution, high-precision, fibre-fed SOPHIE spectrograph, mounted on the 193cm at Observatoire de Haute-Provence, in France \citep{Perruchot08}. The Journal of Observations can be found in Table~\ref{tab:data}. All observations were conducted in HE mode (High Efficiency) where some of the instrumental resolution is sacrificed from 75,000 to 40,000 in favour of a $2.5\times$ greater throughput. We chose this since whilst Kepler-16 is the brightest circumbinary system, it is relatively faint for SOPHIE, with $V \sim 12.0$. 

SOPHIE has two fibres; the first stayed on target, while the other was kept on the sky in order to remove any contribution from the Moon-reflected sunlight. Standard calibrations were made at the start of night as well as roughly every two hours throughout the night to monitor the instrument's zero point. In addition, we observed one of three standards (HD~185144, HD~9407 and HD~89269~A) in HE mode nightly, which we used to track and correct for any long-term instrumental drift following procedures established in \citet{santerne2014sophie} and \citet{Courcol15}. 

Our radial velocities were determined by cross-correlating each spectrum with a K5V mask. These methods are described in \citet{Baranne96}, and \citet{Courcol15}, and have been shown to produce precisions and accuracies of a few meters per second \citep[e.g.][]{Bouchy2013,Hara20}, well below what we typically obtain on this system. As in \citet{Baranne96}, and \citet{pollacco2008wasp}, we correct our data from lunar contamination by first scaling the calculated CCF (cross-correlation function) on fibre A and B (to account for slightly different efficiencies between the two fibres) before subtracting the two CCFs. This is a particularly important procedure for circumbinary planet searches. Most systems observed with SOPHIE are single stars, and the scheduling software informs the observer whether sunlight reflected on the Moon would create a parasitic cross-correlation signal, with a radial velocity that varies predictably with the lunar phase. In such a situation the observation is postponed. In the case of binary observations, ours, the velocity of the primary star keeps changing by $\rm km\,s^{-1}$, meaning we could not practically predict possible lunar contamination at the time of acquisition. 
We also correct our data from the CTI (Charge Transfer Inefficiency) effect following the procedure described in \citet{santerne2012sophie}.

The cross-correlation software produces two key metrics of the shape of the CCF, its FWHM (Full Width at Half Maximum), and its bisector span  \citep[as defined in][]{Queloz:2001}. In addition, we measure the $\rm H\alpha$ stellar activity indicator following \citet{boisse2009stellar}. These are provided in Table~\ref{tab:data}.

Two measurements are immediately excluded from our analysis. On 2017-09-06 (BJD~2,458,003.32008) and 2018-10-06 (BJD~2,458,398.33382), when the fibre was mistakenly placed onto another star. This is obvious from the FWHM we extract from these measurement, and from their radial velocity. They are appropriately flagged in Table~\ref{tab:data}.
In the end, we achieve a mean radial velocity precision of $\sim 10.6~\rm m\,s^{-1}$ on the remaining 141 measurements. 

Radial velocity measurements of the primary have also been obtained with TRES and Keck's HIRES \citep{Doyle:2011vn,Winn:2011wd}. These datasets were not used in this analysis for several reasons. First, the TRES data has a mean precision of $21~ \rm m\,s^{-1}$, which would be insufficient to detect the planet. Second, the HIRES data, despite offering a precision of a few $\rm m\,s^{-1}$, was only taken on a single night and is contaminated by the Rossiter-McLaughlin signal of the eclipse. Finally, by solely using our own SOPHIE data we may produce a near independent detection. In a similar spirit, we do not use any of {\it Kepler}'s or {\it TESS}'s photometric data to conduct our search and parameter estimation.

Finally, \citet{Bender:2012lr} collected near-infrared spectra to reveal the secondary's spectral lines (i.e. observing Kepler-16 as a double-lined spectroscopic binary). Similarly we do not use these measurements in our analysis, although we do use their estimate for the primary mass.

\section{Modelling of the radial velocities}

In order to ascertain our capacity to detect circumbinary planets using radial velocities, as an independent method, we decided to use two different algorithms with different methods for measuring a detection probability. Before describing this, we will detail our procedure to remove outliers. As a reminder, we only use SOPHIE data (see above), where Kepler-16 appears as a single-lined binary and we only observe the displacement of the primary star around the system's barycentre.

\subsection{Outlier removal}\label{sec:outliers}

We searched our 141 data for measurements\footnote{The first series of 14 measurements were obtained from a catalogue containing erroneous proper motions and epochs, which in turn created a $1.5~\rm km\,s^{-1}$ effect in the radial velocity as the Earth's motion was over compensated. This can be corrected easily, but the error remains within the archival data.} coinciding with a primary eclipse, and likely affected by the Rossiter-McLaughlin effect \citep{Rossiter:1924qy,McLaughlin:1924uq,Winn:2011wd,Triaud2018}. Fortunately no measurements needed to be excluded for this reason.

We also realised that a number of measurements were likely taken under adverse conditions. This is apparent from an unusually low signal-to-noise ratio, but also from large values of the bisector span. Typically used as a stellar activity, or a blend indicator \citep[][]{Queloz:2001,Santos:2002fk}, the span of the bisector slope (bisector span, or bis span) effectively informs us that the line shape varied and therefore that the mean of the cross-correlation function is likely affected. We took the mean of the bisector span measurement and removed all measurements in excess of $3\sigma$ away from the mean. Five measurements are excluded this way, all with a bisector span $\gtrsim \pm 100~\rm m\,s^{-1}$. Excluded measurements are reported in the Journal of Observations in Table~\ref{tab:data} with a flag. 

For visual convenience we also exclude one measurement taken on 2018-06-02 (BJD~2,458,271.53928) on account of its very low signal-to-noise and correspondingly large uncertainty, seven times greater than the semi-amplitude of Kepler-16~b. Again, this is reported in Table~\ref{tab:data} with a flag. 

Finally we remove a measurement obtained on 2017-10-30 (BJD~2,458,057.38946). This measurement is $\sim 6\sigma$ away from the best fit model. It is totally unclear why this is the case since its FWHM, bisector span, and $\rm H\alpha$, all appear compatible with other measurements. 

After removing these seven outliers, our analysis is performed on the remaining 134 SOPHIE measurements. The exoplanet detection described just below is done twice, without, and with the seven outlying measurements. Their exclusion did not affect our conclusion but refined our parameters. We also reproduced the following fitting procedures by including the previously existing TRES and HIRES data with no discernible differences.

\begin{figure}
	\includegraphics[width=\columnwidth]{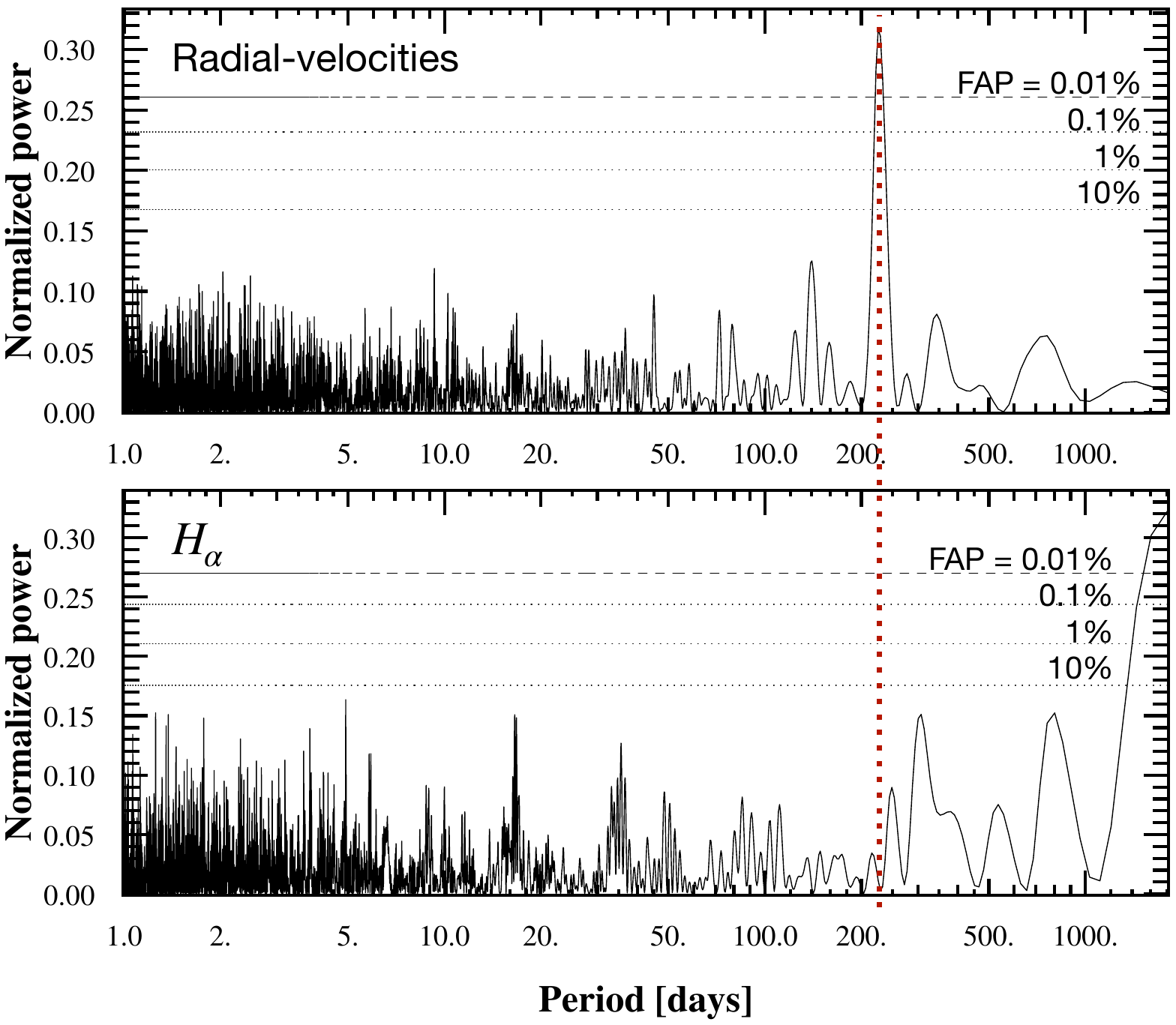}
    \caption{Lomb-Scargle periodogram of Kepler-16's radial velocities (top) and $\rm H\alpha$ (bottom). The radial velocities are shown after removing the binary motion, and a cubic function. The four lines are, from bottom to top, the $10\%$, $1\%$, $0.1\%$ and $0.01\%$ false alarm probabilities. There is a highly significant peak around 230 days (vertical red dotted line) that is present in the radial velocities but not in $\rm H\alpha$. The $\rm H\alpha$ measurements contain significant periodogram power at $\gtrsim 2000~\rm days$, indicating a long-term trend in the chromospheric emission from the primary star.}
    \label{fig:periodogram}
\end{figure}

\subsection{Analysis using the genetic algorithm {\sc Yorbit}}

{\sc Yorbit} is a radial velocity fitting tool used for exoplanet detection, described in \citet{Segransan:2011}. It assumes Keplerian orbits (see Sect~\ref{sec:orbital_elements}). {\sc Yorbit} first performs a Lomb-Scargle periodogram of the observations, which is then used to initiate a genetic algorithm that iterates over the orbital period $P$, the eccentricity $e$, the argument of periastron $\omega$, the semi-amplitude $K$, a reference time $T_0$, and one systemic velocity per dataset $\gamma$ \citep[for conventions, see][]{Hilditch:2001uq}. Once the algorithm has converged on a best solution, the final parameters are estimated by using a least-square fit. The tool has been routinely used to identify small planetary companions successfully \citep[e.g.][]{Mayor:2011fj,Bonfils:2013fk}, and represents a more traditional, and possibly a more recognised way of identifying a new planetary system than the nested sampler we use subsequently (see Sect.~\ref{sec:kima}). However {\sc Yorbit} cannot do a Bayesian model comparison. Instead it computes a False Alarm Probability (FAP) by performing a bootstrap on the data thousands of times and computing for each iteration a Lomb-Scargle periodogram.

In a first instance, {just one Keplerian is adjusted to the SOPHIE data, with {\sc Yorbit} automatically finding the most prominent signal, that of} the secondary star. {Following this step, we remove the secondary's signal and search the resulting residuals with a} periodogram. {This periodogram shows} excess power around 230 day with a $\rm FAP~\sim 0.1\%$ as well as excess power for a signal longer than the range of dates we observed for, which we later associate to a magnetic cycle. {To isolate the signal of Kepler-16~b,} we fit the secondary's Keplerian to the data, alongside a polynomial function, which is used to detrend {that longer signal. Once {\sc Yorbit} has converged, we search the residuals again with a periodogram, which provides} a $\rm FAP~\ll 0.01\%$ (Fig.~\ref{fig:periodogram}), clearly detecting Kepler-16~b as an additional periodic sinusoidal signal. The FAP obtained with a cubic {detrending} function is one order of magnitude better than that obtained with a quadratic function, so we chose the former as our baseline detrending. 

\begin{figure}
	\includegraphics[width=\columnwidth]{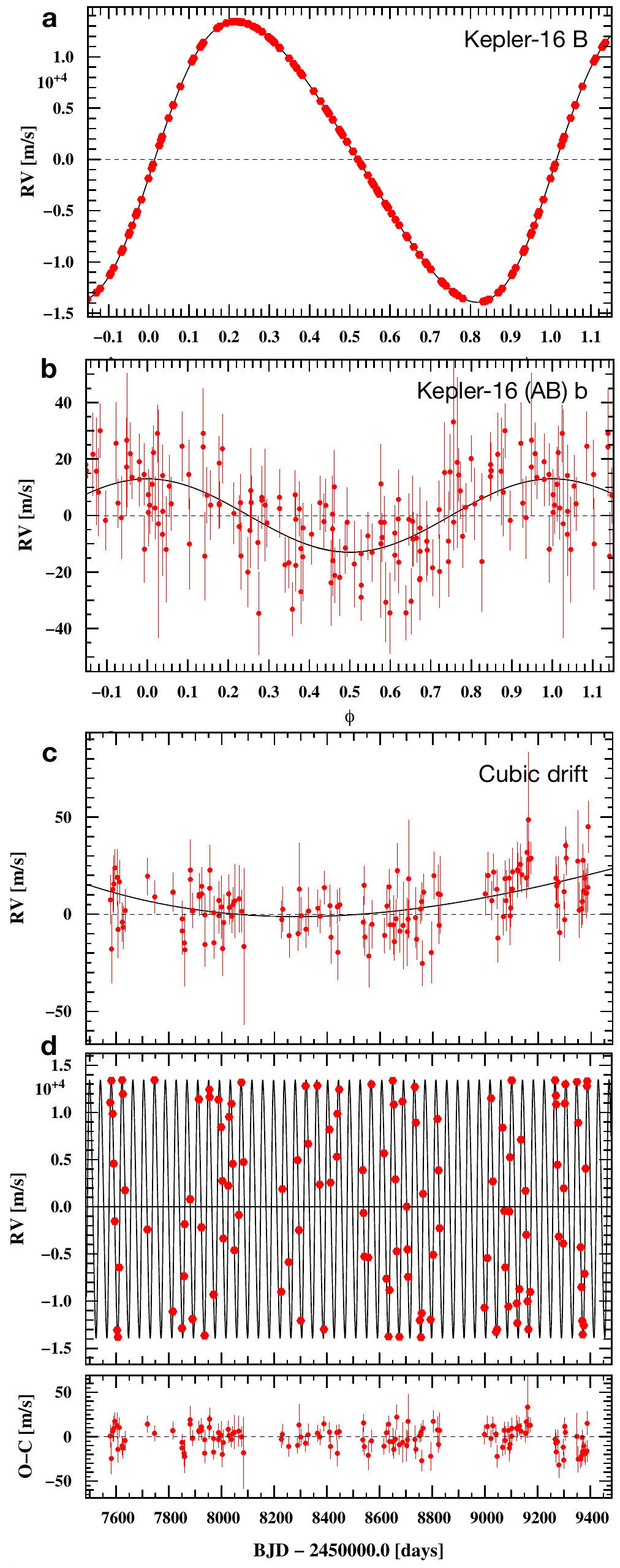}
    \caption{Best fit adjustment to the SOPHIE radial velocity data. {\it a:} Doppler reflex motion caused by the secondary star. {\it b:} Doppler reflex motion caused by the circumbinary planet. {\it c:} Cubic drift associated with a magnetic cycle. {\it d:} radial velocities as a function of time with a binary+planet+cubic function model. Residuals are displayed below.}
    \label{fig:orbit}
\end{figure}

{To obtain results on the system's parameters, we perform a final fit to the data assuming two Keplerians and a cubic detrending function. Results of that fit are} found in Table~\ref{tab:results}. The orbital parameters of the planet are compatible with those produced in \citet{Doyle:2011vn} (see table~\ref{tab:results} and section~\ref{sec:orbital_elements} for a discussion). Our final fit produces a reduced $\chi^2_\nu = 1.17\pm0.14$, implying no additional complexity is needed to explain the data, and supporting our choice for a circular planetary orbit\footnote{Making a fit with a quadratic function we obtain $\chi^2_\nu = 1.47\pm0.15$, for one fewer parameter, justifying our choice for the cubic drift.}. In addition, this shows that we can achieve photon-noise precision on single-lined binary to detect circumbinary planets. The model fit to the data is depicted in Fig.~\ref{fig:orbit}.

Including seven outliers described in Sect.~\ref{sec:outliers}, neither the FAP nor the reduced $\chi^2_\nu$ are significantly affected. 

\begin{table}
	\centering
	\caption{Results of our analysis of the SOPHIE radial velocities only, {after removing outliers, that show} the fit's Jacobi parameters and their derived physical parameters. They are compared to previous results with $1\sigma$ uncertainties provided in the form of the last two significant digits, within brackets. Dates are given in BJD - 2,450,000. We adopt the {\sc Kima} column as our results. }
	\label{tab:results}
	\begin{tabular}{lllll} 
		\hline
		\hline
		\multicolumn{2}{l}{Parameters \& units} & {\sc Yorbit} & {\sc Kima} & Doyle+ \citeyearpar{Doyle:2011vn}\\
		\hline
		\multicolumn{5}{l}{\it binary parameters}\\
		$P_{\rm bin}$        &day             & 41.077779(54) & 41.077772(51) & 41.079220(78)\\
		$T_{0, \rm bin}$     &BJD             & 8558.9640(44) & 7573.0984(47) & --\\
		$K_{1, \rm bin}$     &$\rm m\,s^{-1}$ & 13\,678.2(1.5)  & 13\,678.7(1.5) & --\\
		$e_{\rm bin}$        &--              & 0.15989(11)   & 0.15994(10) & 0.15944(62)\\
		$\omega_{\rm bin}$   &deg             & 263.661(40)   & 263.672(40) & 263.464(27)\\
		\\
		\multicolumn{5}{l}{\it planet parameters}\\
		$P_{\rm pl}$         &day             & 228.3(1.8)  & 226.0(1.7) & 228.776(37)\\
		$T_{0, \rm pl}$      &BJD             & 8532.5(4.4) & 7535(92) & --\\
		$K_{1, \rm pl}$      &$\rm m\,s^{-1}$  & 12.8(1.5)   & 11.8(1.5) & --\\
		$e_{\rm pl}$         &--              & 0 (fixed)   & $<$0.21   & 0.0069(15)\\
		$\omega_{\rm pl}$    &deg             & --          &   231(65)    & $318^{(+10)}_{(-22)}$\\
		\\
		\multicolumn{5}{l}{\it system parameters}\\
		$\gamma$             &$\rm km\,s^{-1}$  & -33.8137(69) & -33.8065(45) & -32.769(35) \\
		$\sigma_{\rm jitter}$ &$\rm m\,s^{-1}$  & --           & 0.070 $^{+1.104}_{-0.067}$ & -- \\
		\hline
		\multicolumn{5}{l}{\it derived parameters}\\
		$M_1$         &$\rm M_\odot$& 0.654(17)$^*$ & 0.654(17)$^*$ & 0.6897(35)\\
		$M_2$         &$\rm M_\odot$& 0.1963(31) & 0.1964(31) & 0.20255(66)\\
		$m_{\rm pl}$  &$\rm M_{Jup}$& 0.345(41) & 0.313(39) & $0.333 (16)$\\
		$a_{\rm bin}$ &$\rm AU$     & 0.2207(18) & 0.2207(17) & $0.22431 (35)$\\
		$a_{\rm pl}$ &$\rm AU$      & 0.6925(67) & 0.6880(58) & $0.7048 (11)$\\
		\hline
	\end{tabular}
	{* adopted from \citet{Bender:2012lr}}
\end{table}

\subsection{Analysis using the diffusive nested sampler {\sc Kima}}\label{sec:kima}
{\sc Kima} is a tool developed by \citet{kima} which fits a sum of Keplerian curves to radial velocity data. It samples from the posterior distribution of Keplerian model parameters using a Diffusive Nested Sampling algorithm by \citet{Brewer2016}. Diffusive Nested sampling allows the sampling of multi-modal distributions, such as those typically found in exoplanetary science and radial velocity data \citep{Brewer2015}, evenly and efficiently. 

{\sc{Kima}} can treat the number of planetary signals ($N_{\rm p}$) present in an RV data set as a free parameter in its fit. Since the tool also calculates the fully marginalised likelihood (evidence) it allows for Bayesian model comparison \citep{Trotta2008} between models with varying $N_{\rm p}$. A measure of preference of one Bayesian model over another can be ascertained by computing the "Bayes Factor" \citep{KassRaftery1995} between the two. The Bayes factor is a ratio of probabilities between the two competing models. Once a value for the Bayes factor has been calculated, we can compare it to the so-called ``Jeffreys' scale" (see \citet{Trotta2008} for more details) to rate the strength of evidence of one model over another.

A more extensive description of our use of {\sc{Kima}} in the context of the BEBOP survey can be found in \citet{Standing2021}.
For the analysis of the Kepler-16 system prior distributions were chosen similarly to those used in \citet{faria2020} with the following notable adaptions. 
We treat the secondary star as a \textit{known object} with tight uniform priors on its orbital parameters. 
A log-uniform distribution was used to describe the periods of any additional signals, from $4 \times P_{\rm bin}$ to $1\times10^4$ days. This inner limit on period is set by the instability limit found in binary star systems \citep{Holman:1999lr}. More details, particularly on the priors we use, can be found in \citet{Standing2021}. {Just like for the previous analysis using {\sc Yorbit}, {\sc Kima} is only deployed on SOPHIE data, and excluding outliers described in Sec.~\ref{sec:outliers}. }

Our {\sc{Kima}} analysis of the Kepler-16 data yields a Bayes Factor $\rm BF > 10000$ in favour of a three Keplerian model (secondary star, planet and cubic drift). Our posterior shows over-densities at orbital periods of $\approx 230$ and $\approx 2000$ days, corresponding to the signal of Kepler-16~b and the cubic drift seen in {\sc{Yorbit}}. We then apply the clustering algorithm {\sc{HDBSCAN}} \citep{McInnes2017} to isolate and extract the resulting planetary orbital parameters, which can be found in table~\ref{tab:results}. 

\subsection{Note on converting fitted parameters to physical values}

To convert our semi-amplitudes into masses for $M_2$ and $m_{\rm pl}$, for the secondary star and planet masses (respectively), we adopt a primary star mass ($M_1$) from \citet{Bender:2012lr}. 
Software written for exoplanetary usage usually assumes that $m_{\rm pl} \ll M_\star$ (including {\sc Yorbit} and {\sc Kima}), however this assumption is no longer valid when comparing $M_2$ to $M_1$, and a circumbinary planet to both. 

First we find $M_2$ iteratively using the mass function, following the procedure described in \citet{Triaud:2013lr}. Then, we estimate $m_{\rm pl}$ from $K_{1,\rm pl}$ by using the combined mass $M_1 + M_2$. This is because whilst we are only measuring the radial velocity signature of the primary star, the gravitational force of the planet acts on the barycentre of the binary.
Had we not done this extra conversion step, we would find significant differences, with erroneous $M_2 = 0.165~\rm M_\odot$ and $m_{\rm pl} = 0.29~\rm M_{Jup}$ for the {\sc Yorbit} results.

Differences between our derived parameters (bottom part of table~\ref{tab:results}) and those from \citet[][]{Doyle:2011vn} are mainly explained by our adoption of the more accurate $M_1$ mass from \citet{Bender:2012lr} rather than using the value from \citet[][]{Doyle:2011vn}.

\subsection{Note on circumbinary planets' orbital elements}\label{sec:orbital_elements}

The main differences between our fitted parameters and those from \citet{Doyle:2011vn} are caused by our parameters being akin to mean parameters (e.g. the mean orbital period), whereas \citet{Doyle:2011vn} provide osculating parameters, which are the parameters the system had at one particular date, and which constantly evolve following three-body dynamics \citep{Mardling:2013}. Also, the planetary signal is significantly more obvious in the {\it Kepler} transiting data than in our radial velocities. Each measurement within a planetary transit over the primary star produces an $\rm SNR = 243$ ($\rm SNR=14$ when over the secondary). This is why {\it Kepler} can derive osculating elements, by solving Newton's equations of motion: the planetary motion is resolved orbit after orbit (transit after transit).  Comparatively, our radial velocity observations have required multiple orbits of the planet to build up a significant detection. We can therefore only measure a mean period, and are justified in using software which can only adjust non-interacting Keplerian functions.

For example, \citet{Doyle:2011vn} provide a highly precise value of $P_{\rm pl} = 228.776 \pm 0.037$ days, yet this osculating period will vary by approximately $\pm 5$ days over a timescale of just years. Our measured period of $P_{\rm pl}=226.0\pm1.7$ days has a much higher error with this uncertainty being a combination of our radial velocities being less constraining on the period, and our assumption of a static orbit. The value we obtain with {\sc Kima} is $1.6\sigma$ compatible with the value found by \citet{Doyle:2011vn}.

With respect to the planet mass, we can derive a value with similar precision to that of \citet{Doyle:2011vn} because whilst the transit signature of the planet is much stronger than its radial velocity signature, the transit signature itself carries very little information about the planet's mass. The photodynamical mass derived in \citet{Doyle:2011vn} is dictated by the eclipse timing variations, which have an amplitude of a couple of minutes and have a precision on the order of tens of seconds, producing an SNR close to our radial velocities.

Finally, to validate our assumption that we cannot measure osculating elements with our current data, we used tools described in \citet{Correia:2005lr} and \citet{Correia_etal_2010} to perform a N-body fit to the radial velocities. This fit finds no improvements in $\chi^2$ indicating that Newtonian effects indeed remain below the detectable threshold. 

\subsection{A magnetic cycle, and constraints on additional planets}

To assess the presence of an external companion causing the additional polynomial signal we notice in our data, we force a two-Keplerian model to fit the data (a circular orbit for the planet and a free-eccentricity orbit for the binary) and analyse the residuals. We find a signal reaching a  $\rm FAP~< 0.01\%$ at orbital periods exceeding the timespan of our data ($\gtrsim 2,000 \rm~days$). Long-term drifts can sometimes be caused by magnetic cycles since stellar spots and faculae tend to suppress convective blue-shift, producing a net change in the apparent velocity of a star \citep{Dravins:1985,Meunier:2010,Dumusque:2011}. We perform a Lomb-Scargle periodogram on the $\rm H\alpha$ activity indicator we extracted from the SOPHIE spectra, and find a $\rm FAP<0.01\%$. We therefore interpret this long-term radial velocity drift as a magnetic cycle, with a timescale longer than the timespan of our observations.

The validity of this interpretation can be tested since duration of stellar magnetic cycles scales with stellar rotation periods \citep[for single star;][]{Suarez:2016}. We measure the primary star's rotation from the $v \sin i_1$ obtained during the Rossiter-McLaughlin effect by \citet{Winn:2011wd} to the primary's stellar radius obtained by \citet{Bender:2012lr} and obtain a primary rotation $P_{\rm rot,1} = 35.68\pm 1.04~\rm days$. 
Following the relation of \citet{Suarez:2016}, with the observed stellar rotation we ought to expect a magnetic cycle on a timescale of 1900 to 2400 days, which is entirely compatible with the $\rm H\alpha$ signal and the long-term radial velocity drift.

\begin{figure}
	\includegraphics[width=\columnwidth]{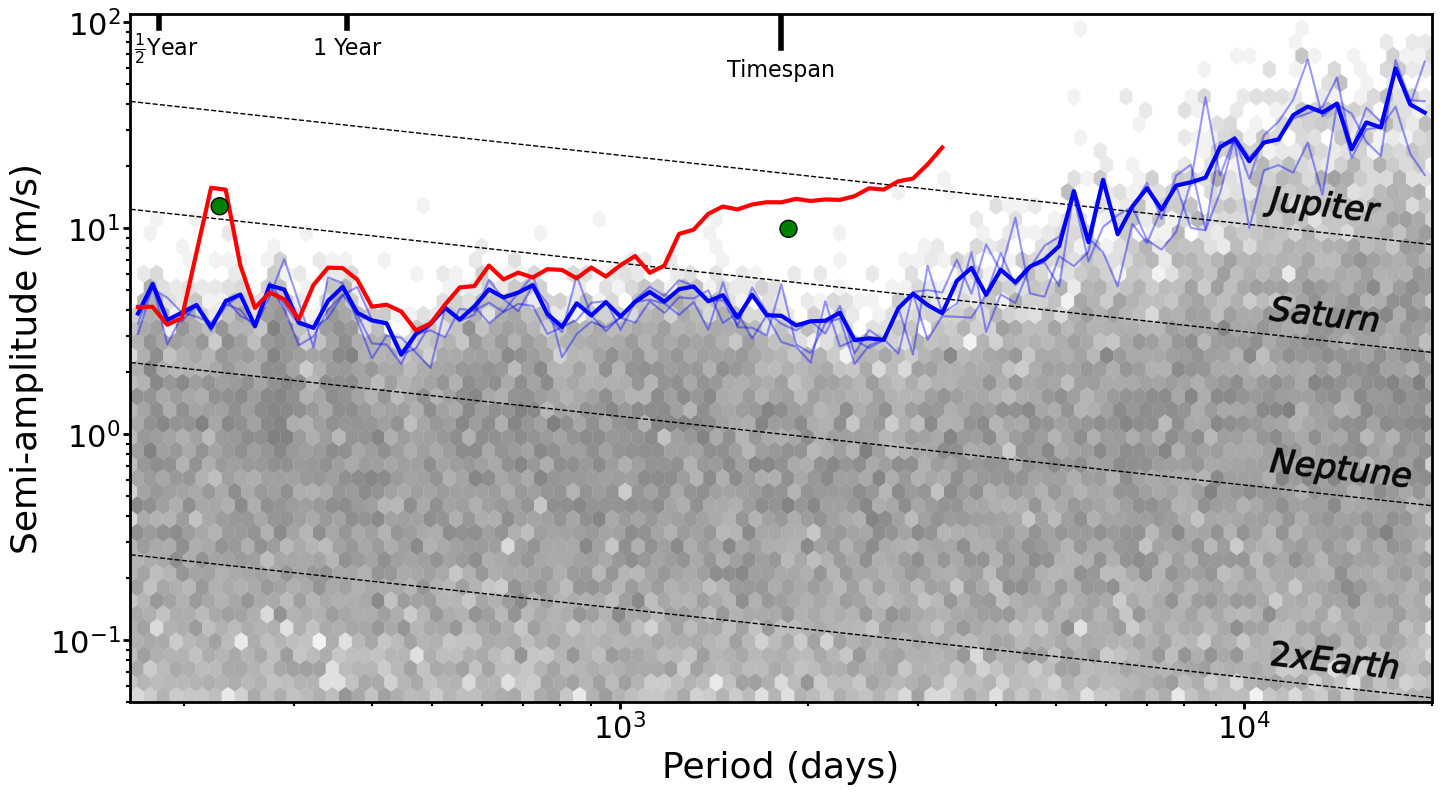}
    \caption{Detection sensitivity to additional planets plotted as semi-amplitude $K_{1,\rm pl}$ as a function of $P_{\rm pl}$. The hexagonal bins depict the density of posterior samples obtained from three separate {\sc Kima} runs applied on the Kepler-16 radial velocity data data after removing two Keplerian signals shown with green dots. The faded blue lines show detection limits calculated for each of the three runs on the system. The solid blue line shows the detection limit calculated from all posterior samples combined. The solid red line is the outline of the posterior that led to the detection of Kepler-16~b and of a long-term trend associated with a magnetic cycle. The green dots represent the two signals removed from the data to compute the blue detection limit.}
    \label{fig:detectionlimit}
\end{figure}

We now use {\sc Kima} as done in \citet{Standing2021} to compute a detection limit on the presence of additional but undetected planetary companions. We first remove the highest likelihood model with two Keplerians (the planet and the long-term drift) from the data, then force $N_{\rm p} = 1$ to obtain a map of all remaining signals that are compatible with the data, but remain formally undetected (the binary is also adjusted at each step). This map is shown as a greyscale density on Fig.~\ref{fig:detectionlimit}. The $99\%$ contour informs us that we are sensitive to companions below the mass of Kepler-16~b up to orbital periods of $\sim3,000$ days. 
This complements \citet{Martin2021}'s work who placed detection limits down to Earth-radius planets but only out to periods of 500 days, re-analysing the {\it Kepler} photometry. 

Overall, a picture is emerging of Kepler-16\,b as a lonely planet, which has implications for the formation and migration of multi-planet systems in the presence of a potentially destabilising binary \citep[e.g.][]{Sutherland:2019}.

\section{Discussion}\label{sec:discussion}

Our data clearly shows an independent detection of the circumbinary planet Kepler-16~b, the first time a circumbinary planet is detected using radial velocities, and the first time a circumbinary planet is detected using ground-based telescopes as well. We re-iterate that our model fits are solely made using radial velocities and completely ignore any {\it Kepler} or other photometric data, to emulate BEBOP's blind search. Importantly our results show we can achieve a precision close to $\rm 1~m\,s^{-1}$ on a planetary signal, with the $1\sigma$ uncertainties on the semi-amplitudes being only just $\rm 1.5~m\,s^{-1}$. This is compatible with the semi-amplitudes of super-Earths and sub-Neptunes \citep[e.g.][]{Mayor:2011fj}. The closest previous detection to a circumbinary planet made from the ground was produced by \citet{Correia:2005lr} on HD~202206, a system comprised of a Sun-like star with an inner companion with mass $m_b\sin i_b = \rm 17.4~M_{Jup}$ and orbital period $P_b = 256~\rm days$, and an outer companion with mass $m_c\sin i_c = \rm 2.44~M_{Jup}$ and orbital period $P_c = 1383~\rm days$. \citet{Benedict:2017} claim an astrometric detection of the system that implies a nearly face-on system, with $m_b = 0.89~\rm M_\odot$ and $m_c = 18~\rm M_{Jup}$, suggesting a circumbinary brown dwarf. However a dynamical analysis produced by \citet{Couetdic2010} imply such a configuration is unstable and therefore unlikely, favouring instead a more edge-on system.

We first detect Kepler-16~b by using a classical approach to planet detection, via periodograms and false alarm probabilities (FAP), but also perform a second analysis using the diffusive nested sampler {\sc Kima}, which allows to perform model comparison and model selection in a fully Bayesian framework. {\sc Kima} will be the method of choice for the remainder of the BEBOP survey \citet{Standing2021}. The parameters for the planetary companion are broadly compatible with those measured at the time of detection by \citet{Doyle:2011vn}, and which have not been revised since. With our current precision it is not possible to determine the eccentricity of the orbit, but we can place an upper limit on it.

\begin{figure}
	\includegraphics[width=\columnwidth]{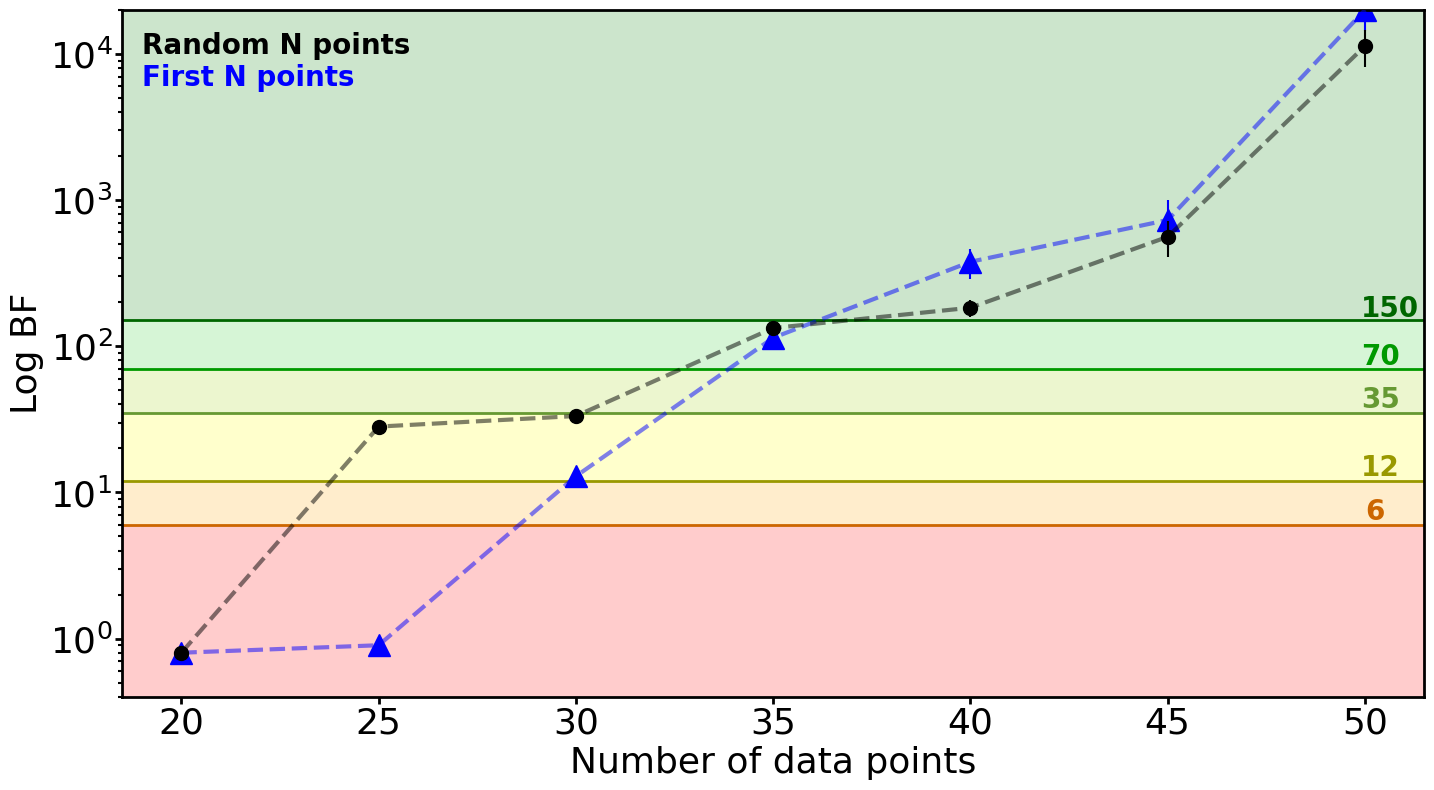}
    \caption{Bayes factors obtained for subsets of data with increasing size. The blue line and triangles indicate the Bayes factors obtained by running {\sc Kima} on the first \textit{N} data points obtained on Kepler-16. The black line and dots show the same but for randomly sampled \textit{N} data points within the first 3 years of data. Coloured regions represent various thresholds corresponding to improvements in evidence in favour of the more complex binary $+$ planet model as per the Jeffreys' scale detailed in \citet{Standing2021}.}
    \label{fig:test}
\end{figure}

Finally we discuss the detectability of circumbinary systems. 
We only take the 20 first measurements we collected and run {\sc Kima} measuring the Bayes Factor to assess the detectability of Kepler-16~b, where individual uncertainties are similar to the semi-amplitude of the signal, $K_{1,\rm pl}$. We repeat the procedure, measuring BF for each increase of five measurements until we reach a sub-sample containing the 50 first radial velocity measurements (they roughly cover four orbital periods of the planet). We reach a BF $>150$ (the formal threshold for detection) with the first 40 measurements. We repeat this procedure but instead randomly select 50 measurements within the first three years of data. We run {\sc Kima} and measure the Bayes Factor for the first 20 epochs of this sequence of 50 random epochs. We then increase from 20 to 25 until reaching 50. We find that the BF~$=150$ threshold is also passed at 40 measurements. We plot our results in Fig.~\ref{fig:test}. The results between both series of tests are broadly consistent, except when we only use 25 measurements, with the Bayes Factor growing log-linearly with increasing number of measurements. Thanks to these tests we conclude that just 40 to 45 measurements would have been to formally detect Kepler-16\,b with radial velocities.

\section*{Acknowledgements}

First, we would like to thank the staff, in particular the night assistants, at the Observatoire de Haute-Provence for their dedication and hard work, particularly during the COVID pandemic. 
The data were obtained first as part of a DDT graciously awarded by the OHP director (Prog.ID 16.DISC.TRIA), and then by a series of allocations through the French PNP (Prog.IDs 16B.PNP.HEB2, 17A.PNP.SANT, 17B.PNP.SAN2, 18A.PNP.SANT, 18B.PNP.SAN1, 19A.PNP.SANT). The French group acknowledges financial support from the French Programme National de Plan\'etologie (PNP, INSU).
This research received funding from the European Research Council (ERC) under the European Union's Horizon 2020 research and innovation programme (grant agreement n$^\circ$ 803193/BEBOP) and from the Leverhulme Trust (research project grant n$^\circ$ RPG-2018-418). EW acknowledges support from the ERC Consolidator Grant funding scheme (grant agreement n$^\circ$772293/ASTROCHRONOMETRY). 
Support for this work was provided by NASA through the NASA Hubble Fellowship grant number HF2-51464 awarded by the Space Telescope Science Institute, which is operated by the Association of Universities for Research in Astronomy, Inc., for NASA, under contract NAS5-26555.
P.C. thanks the LSSTC Data Science Fellowship Program, which is funded by LSSTC, NSF Cybertraining Grant \#1829740, the Brinson Foundation, and the Moore Foundation; her participation in the program has benefited this work. 
SH acknowledges CNES funding through the grant 837319. ACC acknowledges support from
the Science and Technology Facilities Council (STFC) consolidated
grant number ST/R000824/1.
{ACMC acknowledges support by CFisUC projects (UIDB/04564/2020 and UIDP/04564/2020), GRAVITY (PTDC/FIS-AST/7002/2020), ENGAGE SKA (POCI-01-0145-FEDER-022217), and PHOBOS (POCI-01-0145-FEDER-029932), funded by COMPETE 2020 and FCT, Portugal.}

\section*{Data Availability}

The radial velocity data are included in the appendix of this paper. The reduced spectra are available at the SOPHIE archive: \href{http://atlas.obs-hp.fr/sophie/}{http://atlas.obs-hp.fr/sophie/}. Data obtained under Prog.IDs 16.DISC.TRIA, 16B.PNP.HEB2, 17A.PNP.SANT, 17B.PNP.SAN2, 18A.PNP.SANT, 18B.PNP.SAN1, and 19A.PNP.SANT.



\bibliographystyle{mnras}
\bibliography{Kep16} 




\appendix

\section{Journal of Observations}


\onecolumn
\begin{longtable}[1]{cllllllll}\label{tab:data}\\
	\caption{Journal of Observations containing our SOPHIE data. Flags indicate whether the measurement are excluded from our fiducial analysis with the following reason: W, wrong star; B, bisector outlier; U, high uncertainty; O, other. Dates are given in BJD - 2,400,000. $V_{\rm rad}$ are the measured radial velocities with their uncertainties $\sigma_{V_{\rm rad}}$. FWHM is the Full With at Half Maximum of the Gaussian fitted to the cross correlation function, and {\it contrast} is its amplitude. Bis. span is the span of the bisector slope. Uncertainties on FWHM and bis. span are $2\times\sigma_{V_{\rm rad}}$. $H_\alpha$ is the equivalent width of $H_\alpha$, and its uncertainty $\sigma_{H_\alpha}$.}\\
	\hline
	\hline
    flag&BJD-2,400,000 &$V_{\rm rad}$&$\sigma_{V_{\rm rad}}$&FWHM&contrast&bis. span&$H_\alpha$&$\sigma_{H_\alpha}$\\
    &[days]&[$\rm km\,s^{-1}$]&[$\rm km\,s^{-1}$]&[$\rm km\,s^{-1}$]& & [$\rm km\,s^{-1}$] &&\\
	\hline
	\endfirsthead
	\hline
	\hline
    flag&BJD&$V_{\rm rad}$&$\sigma_{V_{\rm rad}}$&FWHM&contrast&bis. span&$H_\alpha$&$\sigma_{H_\alpha}$\\
	\hline
    \endhead
    \hline
    \multicolumn{5}{l}{Table continues next page...}\\
    \hline
    \endfoot
    \hline
    \endlastfoot 
&57578.42404&-22.7567&0.0126&10.4138&22.7274&-0.0233&0.2119&0.0022\\
B&57581.57256&-20.3465&0.0444&10.4924&13.0881&-0.1227&0.2184&0.0055\\
&57582.52743&-20.4322&0.0176&10.3201&27.2220&-0.0199&0.2046&0.0036\\
&57587.43166&-23.9495&0.0113&10.2935&25.8863&0.0192&0.2031&0.0021\\
&57591.44122&-29.2376&0.0146&10.2883&26.1452&0.0538&0.2117&0.0029\\
&57595.47979&-35.3428&0.0096&10.3340&29.0424&0.0125&0.2039&0.0020\\
&57604.43905&-46.8841&0.0145&10.4605&25.3376&-0.0054&0.2147&0.0028\\
&57607.38886&-47.6294&0.0145&10.4123&28.2874&0.0037&0.2135&0.0030\\
&57612.47728&-40.2406&0.0110&10.3402&28.9602&0.0079&0.2116&0.0023\\
&57623.37736&-20.3825&0.0100&10.2928&29.0664&-0.0093&0.2032&0.0020\\
&57626.38491&-21.8607&0.0110&10.3395&28.9280&0.0329&0.2063&0.0022\\
&57634.40704&-32.0631&0.0109&10.3925&28.9876&-0.067&0.2092&0.0023\\
&57719.26505&-36.2183&0.0095&10.3290&28.9030&0.0201&0.1966&0.0018\\
&57746.26923&-20.3795&0.0084&10.3404&29.0733&-0.0067&0.2122&0.0016\\
&57815.65363&-44.9376&0.0098&10.3132&27.1330&-0.0004&0.2095&0.0018\\
&57815.67731&-44.8993&0.0083&10.3249&29.1040&0.0086&0.2080&0.0016\\
&57850.59947&-46.6829&0.0091&10.3250&27.0881&0.0033&0.2069&0.0017\\
&57850.62308&-46.7070&0.0119&10.2789&28.0442&-0.0021&0.2067&0.0022\\
&57858.58652&-41.1658&0.0146&10.2378&23.9695&-0.0255&0.1882&0.0025\\
&57860.61247&-35.6563&0.0186&10.2443&23.5034&0.0100&0.1941&0.0032\\
&57881.49587&-33.0003&0.0175&10.2694&22.2896&-0.0211&0.2216&0.0030\\
&57881.51944&-33.0316&0.0158&10.2409&25.8066&0.0089&0.2109&0.0029\\
&57890.53416&-45.6792&0.0089&10.2779&27.3843&0.0160&0.1951&0.0016\\
&57890.57816&-45.7231&0.0081&10.2384&27.7725&-0.0009&0.1910&0.0015\\
&57914.55102&-22.4041&0.0110&10.2004&25.3872&-0.0078&0.2151&0.0020\\
&57914.57472&-22.4272&0.0084&10.2139&27.9878&0.0084&0.2134&0.0016\\
&57924.49590&-35.9753&0.0099&10.2853&27.2617&0.0267&0.1996&0.0018\\
&57924.52118&-36.0181&0.0103&10.2534&28.8795&0.0171&0.2012&0.0020\\
&57936.48767&-47.4623&0.0106&10.2734&26.9318&0.0141&0.1932&0.0020\\
&57936.51133&-47.4660&0.0113&10.3536&28.5829&-0.0057&0.1917&0.0021\\
&57954.37721&-21.3857&0.0134&10.2930&25.3470&0.0145&0.2031&0.0024\\
&57955.36873&-22.1650&0.0129&10.2899&25.3384&-0.0119&0.2043&0.0022\\
&57970.44413&-43.0893&0.0085&10.2438&28.2732&0.0268&0.1914&0.0016\\
&57970.46806&-43.1358&0.0082&10.2520&28.9462&0.0077&0.1877&0.0015\\
&57989.42736&-22.4627&0.0097&10.2269&27.1918&0.0127&0.1965&0.0017\\
&57989.45189&-22.4383&0.0091&10.2555&29.0876&-0.0119&0.1930&0.0017\\
&57999.37737&-25.3762&0.0160&10.2317&23.9715&-0.0013&0.1971&0.0027\\
W&58003.32008&-3.1611&0.0199&7.5424&24.2589&0.0161&0.2016&0.0045\\
&58003.42095&-31.0632&0.0141&10.1974&23.2722&-0.0428&0.1954&0.0023\\
&58007.42352&-37.1754&0.0142&10.3020&28.2437&0.0059&0.1924&0.0027\\
&58026.32236&-31.5767&0.0100&10.2828&26.6464&0.0082&0.2004&0.0018\\
&58029.34747&-24.2823&0.0086&10.2444&31.1481&-0.0078&0.1962&0.0018\\
B&58034.33565&-20.4064&0.0114&10.2145&31.4902&-0.0954&0.1925&0.0024\\
&58038.28757&-22.8921&0.0085&10.2171&29.0859&-0.0010&0.1947&0.0016\\
&58043.30107&-29.2555&0.0088&10.1869&27.6237&0.0084&0.1888&0.0016\\
&58049.33656&-38.4188&0.0190&10.3105&23.6222&-0.0004&0.2151&0.0033\\
O&58057.38946&-47.4170&0.0127&10.3516&27.6078&0.0028&0.1924&0.0022\\
&58066.32858&-34.6758&0.0173&10.2573&23.9592&0.0012&0.2000&0.0029\\
&58076.24290&-20.6229&0.0109&10.2381&26.3816&0.0227&0.1993&0.0020\\
&58084.24172&-29.0736&0.0402&10.3369&17.3316&0.0183&0.2143&0.0061\\
&58227.56803&-42.8215&0.0095&10.3294&29.2382&0.0159&0.1975&0.0018\\
&58231.59284&-31.9304&0.0115&10.3272&28.2728&-0.0126&0.1911&0.0022\\
&58255.55341&-39.6783&0.0111&10.2862&26.8276&-0.0023&0.1972&0.0020\\
U&58271.53928&-35.2297&0.0775&10.4414&14.1731&0.0514&0.2170&0.0113\\
&58289.48662&-28.8597&0.0086&10.2324&27.6608&0.0124&0.1940&0.0016\\
&58294.41403&-36.2932&0.0238&10.1405&19.6541&-0.0463&0.1980&0.0038\\
&58301.52589&-45.8824&0.0094&10.2594&27.0071&-0.0092&0.1856&0.0017\\
&58319.49236&-21.0086&0.0095&10.232&27.3731&-0.0159&0.1864&0.0017\\
&58329.35035&-27.1417&0.0123&10.1726&26.8217&0.0135&0.1895&0.0024\\
&58364.49871&-20.9710&0.0095&10.2804&29.4757&0.0077&0.1917&0.0018\\
&58373.38714&-31.4460&0.0085&10.2349&28.9319&-0.0064&0.1959&0.0017\\
&58389.28389&-46.7948&0.0085&10.2759&28.4801&-0.0032&0.1936&0.0017\\
W&58398.33382&2.0217&0.0162&7.6074&33.8799&-0.0066&9999.99&9999.99\\
&58410.34703&-25.6250&0.0107&10.2377&30.7540&0.0084&0.1970&0.0023\\
&58414.31942&-31.2461&0.0135&10.2164&26.4540&-0.0110&0.2059&0.0026\\
&58438.23468&-28.5063&0.0086&10.2149&28.7501&-0.0001&0.2048&0.0017\\
&58440.32791&-23.9560&0.0141&10.4258&30.7245&0.0284&0.2063&0.0030\\
&58447.28753&-21.3746&0.0085&10.2282&30.6147&-0.0060&0.1972&0.0018\\
&58536.69186&-29.9300&0.0131&10.2127&26.6076&-0.0381&0.1960&0.0026\\
&58539.69622&-34.4673&0.0101&10.3190&28.0285&-0.0214&0.1959&0.002\\
&58542.70128&-39.0813&0.0096&10.2536&28.4771&-0.0302&0.2042&0.0019\\
&58557.68116&-39.1923&0.0161&10.2757&27.2871&0.0405&0.2019&0.0033\\
&58569.61968&-20.8305&0.0122&10.2296&27.7821&-0.0081&0.2002&0.0024\\
&58617.59111&-28.1421&0.0091&10.2460&28.7484&-0.0038&0.2015&0.0018\\
&58626.46789&-41.4286&0.0107&10.2794&25.8409&0.0075&0.2010&0.0020\\
&58634.53897&-47.5651&0.0093&10.2481&26.5013&-0.0039&0.2018&0.0018\\
&58638.43325&-42.6407&0.0126&10.2794&28.4741&-0.0223&0.2039&0.0026\\
&58650.53936&-20.4374&0.0152&10.1724&26.3894&0.0132&0.2030&0.0031\\
&58654.49571&-22.9659&0.0097&10.2282&28.3680&-0.0006&0.1971&0.0019\\
&58660.56745&-30.9076&0.0102&10.2163&28.9174&-0.0113&0.1984&0.0022\\
&58665.57318&-38.5290&0.0140&10.2965&27.5442&0.0455&0.1981&0.0028\\
&58675.57952&-47.5981&0.0109&10.3365&27.6547&0.0017&0.1965&0.0022\\
&58687.61864&-22.6485&0.0225&10.3376&23.7336&0.0388&0.2054&0.0043\\
&58703.54677&-33.7950&0.0175&10.3134&26.0287&0.0192&0.2010&0.0036\\
&58706.50803&-38.3266&0.0100&10.2295&28.0761&-0.0120&0.1982&0.0020\\
&58708.50232&-41.2314&0.0301&10.2930&16.0707&-0.0426&0.1994&0.0047\\
B&58732.43751&-20.3947&0.0302&10.2203&25.1441&-0.1346&0.2322&0.0062\\
&58734.44238&-21.1078&0.0155&10.2528&27.1499&-0.0207&0.2124&0.0031\\
&58738.38163&-24.8845&0.0105&10.1784&27.7775&-0.0244&0.2003&0.0021\\
&58753.33846&-45.8380&0.0075&10.2737&28.6564&-0.0095&0.202&0.0016\\
&58757.43639&-47.6475&0.0099&10.3475&28.4233&0.0141&0.2001&0.0020\\
&58760.37181&-45.0829&0.0118&10.2839&28.1836&0.0119&0.1973&0.0024\\
&58765.41646&-32.4426&0.0094&10.2947&28.2387&0.0005&0.2029&0.0019\\
&58794.31837&-45.7701&0.0157&10.3120&26.8288&-0.0071&0.2012&0.0031\\
&58804.24157&-38.9032&0.0122&10.3048&27.9886&0.0424&0.2025&0.0024\\
&58820.22529&-24.4962&0.0149&10.3393&26.6102&-0.0071&0.2153&0.0030\\
&58824.24004&-29.9520&0.0094&10.2552&28.2000&0.0245&0.2090&0.0019\\
&58828.27635&-36.0832&0.0198&10.4292&23.1762&-0.0434&0.2019&0.0034\\
B&58919.69124&-47.3685&0.0208&10.3504&24.6286&-0.1229&0.2077&0.0041\\
&58998.56140&-44.5035&0.0109&10.3115&28.8681&0.0086&0.1939&0.0022\\
&59009.49993&-39.2485&0.0122&10.2756&27.7224&-0.0107&0.2039&0.0025\\
&59023.55894&-22.3117&0.0099&10.2759&28.5419&-0.0080&0.1996&0.0019\\
&59030.43413&-31.1164&0.0095&10.2358&28.4005&0.0120&0.1979&0.0019\\
&59042.42642&-47.0737&0.0101&10.2934&28.2747&0.0043&0.2012&0.0020\\
&59046.54366&-46.7980&0.0123&10.2777&28.0726&-0.0441&0.1967&0.0026\\
&59067.42681&-25.4159&0.0097&10.2439&28.0315&0.0085&0.1974&0.0019\\
&59073.55426&-34.2494&0.0093&10.3336&28.3985&0.0340&0.2080&0.0019\\
&59077.48604&-40.2255&0.0157&10.2866&27.4415&0.0209&0.2072&0.0034\\
&59089.39809&-44.3841&0.0099&10.3046&28.1461&0.0012&0.2066&0.0020\\
&59093.41074&-34.3069&0.0151&10.2234&26.0449&-0.0083&0.2102&0.0031\\
&59095.45922&-28.5598&0.0098&10.1601&27.3818&0.0041&0.1988&0.0020\\
&59101.49898&-20.3988&0.0089&10.2738&28.4152&-0.006&0.2009&0.0017\\
&59102.51980&-20.4328&0.0122&10.5231&27.0339&0.0343&0.1998&0.0022\\
&59121.39686&-44.0417&0.0142&10.2192&24.6636&-0.0090&0.2162&0.0028\\
&59123.35979&-46.1410&0.0093&10.2626&27.3614&0.0051&0.2093&0.0018\\
&59131.39916&-42.5350&0.0104&10.2861&28.3114&0.0065&0.2073&0.0021\\
&59137.28463&-26.7021&0.0091&10.2698&28.7466&0.0254&0.2025&0.0018\\
&59154.32126&-32.1154&0.0102&10.3170&26.7867&-0.0023&0.2072&0.0019\\
&59157.36523&-36.7752&0.0177&10.2287&25.1757&0.0102&0.2177&0.0035\\
&59162.32683&-43.8233&0.0349&10.6452&24.2053&-0.0007&0.2148&0.0072\\
&59165.27801&-46.8031&0.0093&10.3282&28.3817&0.0336&0.2066&0.0019\\
&59172.32740&-42.8424&0.0082&10.4652&28.6099&-0.0162&0.2052&0.0016\\
B&59181.27395&-21.7546&0.0168&10.3768&31.0919&-0.1028&0.1955&0.0039\\
&59266.69524&-20.3963&0.0109&10.3016&28.3442&0.0245&0.2087&0.0022\\
&59269.69460&-22.0169&0.0109&10.3060&28.0074&0.0154&0.2067&0.0022\\
&59270.67941&-22.9506&0.0118&10.2768&26.9761&0.0561&0.2142&0.0024\\
&59275.70395&-29.3537&0.0133&10.2461&26.0009&-0.0098&0.2156&0.0027\\
&59280.69715&-36.9685&0.0146&10.2385&27.3492&0.0137&0.2127&0.0030\\
&59297.61346&-37.7124&0.0090&10.3029&28.4720&0.0267&0.1983&0.0018\\
&59299.64512&-31.8527&0.0094&10.3205&28.1204&-0.0358&0.202&0.0019\\
&59303.60959&-22.8602&0.0096&10.2401&27.2753&0.0383&0.1987&0.0019\\
&59305.61783&-20.8265&0.0088&10.2858&28.0542&0.0050&0.2126&0.0018\\
&59349.50200&-20.5659&0.0262&10.4567&25.5230&0.0576&0.2089&0.0054\\
&59354.55499&-24.8984&0.0143&10.2426&28.2716&0.0149&0.2148&0.0031\\
&59363.59631&-38.0943&0.0097&10.2586&27.9776&0.0149&0.2057&0.0020\\
&59366.52159&-42.3251&0.0116&10.2845&28.0478&0.0281&0.2116&0.0024\\
&59369.57437&-45.8980&0.0086&10.3087&28.8846&-0.0363&0.2094&0.0018\\
&59371.53106&-47.3490&0.0106&10.3614&28.6411&-0.0433&0.2137&0.0022\\
&59375.55836&-46.3745&0.0088&10.2746&28.9782&0.0130&0.2074&0.0018\\
&59378.56443&-40.9102&0.0100&10.3196&28.9479&-0.0039&0.2053&0.0021\\
&59382.54493&-29.7562&0.0129&10.2722&28.2695&0.0240&0.2016&0.0027\\
&59387.49783&-21.0110&0.0123&10.2471&26.8552&0.0086&0.2108&0.0024\\
&59388.44241&-20.4912&0.0133&10.2164&26.7626&0.0178&0.2056&0.0027\\
\end{longtable}


\bsp	
\label{lastpage}
\end{document}